\documentclass[twocolumn,showpacs,amsmath,amssymb,prb,superscriptaddress]{revtex4}
\usepackage{graphicx}
\usepackage{bm}
\begin{document}

\title{Screening effects in  Coulomb frustrated phase separation }

\author{C.  Ortix}
\affiliation{Dipartimento di Fisica, Universit\`{a} del Salento and INFN Sezione di Lecce, Via per
  Arnesano, 73100 Lecce, Italy. }
\author{J.  Lorenzana} 
\affiliation{SMC-Istituto Nazionale di Fisica della
Materia and Dipartimento di Fisica, Universit\`a di Roma 
``La Sapienza'', P.  Aldo Moro 2, 00185 Roma, Italy.}
\affiliation{ISC-Consiglio Nazionale delle Ricerche, Via dei Taurini 19, 00185 Roma, Italy.} 
\author{M.  Beccaria}
\affiliation{Dipartimento di Fisica, Universit\`{a} del Salento and INFN Sezione di Lecce, Via per
  Arnesano, 73100 Lecce, Italy. }
\author{C.  Di Castro} 
\affiliation{SMC-Istituto Nazionale di Fisica della
Materia and Dipartimento di Fisica, Universit\`a di Roma 
``La Sapienza'', P.  Aldo Moro 2, 00185 Roma, Italy.}

\date{\today}

\begin{abstract}
We solve a model of phase separation among two competing phases  
frustrated by the long-range Coulomb interaction in two  and three
dimensions (2D / 3D) taking into account finite
compressibility effects. In the limit of strong frustration 
in 2D, we recover the results of R. Jamei,
S. Kivelson, and B. Spivak, Phys.\ Rev.\ Lett. {\bf 94},  056805
(2005) and the system always breaks into domains in a narrow range of
densities, no matter how big is the frustration. 
For weak frustration in 2D and for arbitrary frustration in 3D 
the finite compressibility of the phases is shown to play a
fundamental role. Our results clarify the different role of screening
in 2D and 3D systems. 
We discuss the thermodynamic stability of the system near the
transition to the phase separated state and the possibility to observe
it in real systems. 
\end{abstract}
\pacs{71.10.Hf, 64.75.+g, 71.10.Ca}
\maketitle
\begin{section}{Introduction}
In the presence of long-range forces the 
tendency towards phase separation among two competing phases may 
lead to self-stabilized domain patterns of finite
size \cite{kit46,nag83,cplor93,seu95,sci04}.
A prominent example is provided by  
charged  systems  compensated by a rigid background.
The Coulomb energy cost grows 
faster than the size of the system and precludes macroscopic phase
separation.  Therefore one finds either
uniform phases or  phases with domains of one phase hosted by the
other one \cite{nag83,nag98,cplor93,low94,cas95b,lor01I,lor01II,lor02,mur02,jam05,ort06}.
In the mixed state  the finite Coulomb energy arising from the mismatch
between the local electronic charge $-e n_{i}$ and the fixed background
charge $ e \overline n$, competes with the surface energy to determine
the typical size of the domains.

Experiments in layered materials like cuprates\cite{pan01,mce03,lan02},  
ruthanates\cite{bor07}, manganite thin films\cite{zha02}   
and in the two-dimensional electron gas \cite{lil99,ila01}
have fueled interest in the two-dimensional version 
of this problem \cite{jam05,hon05,ort06}.
In particular the recent discovery of pronounced anisotropies in
transport properties \cite{lil99,bor07}  are consistent with the
proposal of exotic electronic liquid phases \cite{kiv98} analogous to
the intermediate order states of liquid crystals \cite{cha95}.

 Jamei, Kivelson and Spivak \cite{jam05} have analyzed a
  model in which the energies of two infinitely 
 compressible uniform phases cross each other at a certain density $n_c$. 
They have shown that in 2D,  sufficiently close to $n_{c}$, 
the system always breaks into domains for any level 
of frustration due to the long-range Coulomb (LRC) interaction.
They have focused on the universal aspects arising for large frustration 
where the compressibility turns out to be negligible. On the other hand it has been argued that
  screening, driven by finite compressibility, 
 plays a fundamental role on Coulomb frustrated phase
  separation \cite{lor01I,ort06}. It is thus of interest to analyze its
  effect for the full range of frustration and, in this respect, to
  clarify the difference between the two-dimensional (2D) and
  three-dimensional (3D) case.

In this work we solve a  model of Coulomb frustrated phase separation, similar to the one of
Ref.~\onlinecite{jam05} but augmented to  
take  into account the finite compressibility of the phases (Sec.~\ref{sec:model}). This allows us to
derive the phase diagram in 2D from the limit of strong frustration
down to the limit of zero frustration where phase separation is
determined by Maxwell construction (MC) and the compressibility of the
phases, by sure, cannot be neglected. The interplay between the screening and
 the size of inhomogeneities is discussed in two-dimensional systems
 (Sec.~\ref{sec:2d}) and three-dimensional systems
 (Sec.~\ref{sec:3d}). Both cases are compared at the end of
 Sec.~\ref{sec:3d}. We show that for not so large 
frustration in 2D and practically everywhere in 3D the physics of the
mixed state can be captured in a simple approximation where the density
variation inside the inhomogeneities is neglected.  
We argue that the strongly frustrated 2D mixed state regime is
experimentally hardly accessible and characterized by a large negative
electronic compressibility  that can lead to a volume collapse of the
lattice.
The model presented here can be solved exactly and confirms the
conclusions of previous approximated treatments. In particular 
it illustrates the general rule found in Refs.~\onlinecite{lor01I,lor02} 
that for generic value of the parameters inhomogeneities can not have 
all linear dimensions larger that the 3D screening length.

\end{section}
\begin{section}{Model and General Solution}
\label{sec:model}
We consider a ferromagnetic Ising model 
linearly coupled to a charged
fluid as a generic model of a first order density driven
phase transition with Coulomb interactions.  
The Hamiltonian reads:
\begin{eqnarray}
{\cal  H}&=&-J \sum_{<i j >}s_{i}s_{j}-\frac{\Delta\mu_c}2
\sum_{i} s_{i}\left(n_{i}-n_{c}\right) 
\label{eq:model} \\
& &  +\frac{1}{2 \kappa}\sum_{i}\left(n_{i}-n_{c}\right)^{2}
 +\frac{Q^2}{2}\sum_{i j}\left(n_{i}-\overline{n}\right) \dfrac{1}{r_{i j}} \left(n_{j}-\overline{n}\right)\nonumber
\end{eqnarray} 
where $s_i=\pm 1$, the index $i$ runs over the sites of a
hypercubic lattice of dimension $D=2,3$ and $\overline{n}$ is the average charge.
Instead of the present Ising term one could consider, as in Ref.~\onlinecite{jam05} a
double-well potential in the very large barrier limit. In solid state
systems the strength of the Coulomb interaction will be given by
$Q^2=e^2/\epsilon_0$ with $\epsilon_0$ the static dielectric
constant of the environment ({\it i.e.} excluding the mobile electrons).

  For the two uniform phases ($n_{i}=\overline {n}$), the energy as a function
  of density   consists of two parabolas that cross at
$n_c$ with chemical potential difference $\Delta\mu_c$, minima at 
 $\overline{n}=n_c\pm \delta n^0/2$ and $\delta n^0\equiv \kappa
 \Delta\mu_c$  as illustrated by the thick lines in
  Fig.~\ref{energylambda}. For $Q=0$, according to the ordinary Maxwell construction, 
the system phase separates when the average density $\overline{n}$ 
is in a window of width  $\delta n^0$ around the crossing density. 
The energy of this solution is depicted with the dot-dashed horizontal line in Fig.~\ref{energylambda}.

\begin{figure}[tbp]
\includegraphics[width=8 cm]{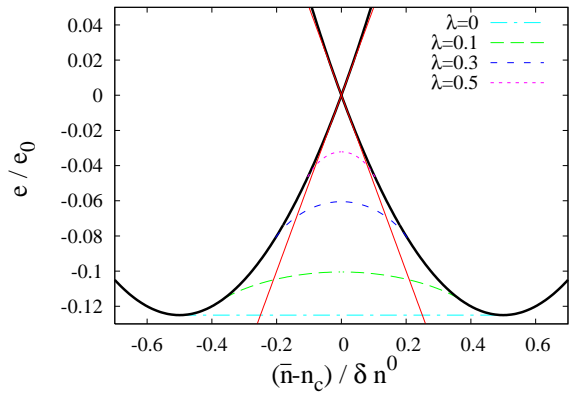}
\caption{(Color online) Behavior of the free energy density for the uniform phases
  with $s=\pm 1$ in the present model (thick lines) compared with the
  $\kappa\rightarrow \infty$ model discussed in Ref.\onlinecite{jam05} 
  (thin full  lines). The dashed lines are the energy density of the mixed state
  solutions in 2D systems for different values of the frustrating
  parameter defined in Sec.~\ref{sec:2d}. The energy density is
  measured in unit of the characteristic phase separation energy gain
  $e_0\equiv (\delta n^{0})^{2} / \kappa=2\pi Q^2  (\delta n^{0})^{2} l_s$. The dot-dashed line
  corresponds to the Maxwell construction ($\lambda=0$).} 
\label{energylambda}
\end{figure}

 We are assuming that the two uniform phases have the same 
compressibility to make the problem solvable. A related model in 3D
has been considered in Ref.~\onlinecite{lor01I}. 
The 2D  case in which one of the phases is 
incompressible  has been analyzed in
Ref.~\onlinecite{ort06},  whereas Ref.~\onlinecite{jam05} corresponds to 
the case in which both phases are infinitely compressible and Maxwell
construction is not defined 
($\kappa\rightarrow \infty$, thin full lines in Fig.~\ref{energylambda}).

Interfaces of the Ising order parameter are sharp  by construction  
with a surface tension given by $\sigma= 2 J/a^{D-1}$ where $a$ is the lattice
constant. The linear term in $\Delta\mu_c$ expresses that 
the homogeneous phase with
$s_i=1$ ($\uparrow$) is favored at higher densities, and the phase with 
$s_i=-1$ ($\downarrow$) at lower densities.    

We solve the problem in the limit in which domains are much larger
than the lattice constant and replace the site index by a continuum
variable, {\it i.e.} $s({\bf r})$,  $n({\bf r})$ in a volume $V$. 
Hereafter we take $a\equiv 1$. For each mixed state, the geometry of the domains
defines  $s({\bf r})$ and its Fourier transform $s({\bf q})$. Writing the energy in the
Fourier space, the ${\bf q}=0$ component of the Coulomb term is canceled. 
At ${\bf q} \neq 0$, upon minimizing with respect to the charge distribution,
one gets:
\begin{equation}
n({\bf q})\left(\kappa^{-1}+2^{D-1}\pi Q^2 / |{\bf q}|^{D-1}
\right)=\frac{\Delta\mu_c}2 s({\bf q}) 
\label{eq:nq}
\end{equation}

This equation has a simple physical interpretation by noticing that for a fixed domain configuration
$$\mu\left({\bf r}\right)\equiv -\dfrac{\Delta\mu_{c}}{2} s\left({\bf r}\right) + \kappa^{-1} \left[n\left({\bf r}\right)-n_c\right]$$ 
represents the local chemical potential and the electrostatic potential can be put as:
$$v\left({\bf r}\right)=\int d{\bf r'}  \phi\left({\bf r-r'}\right) n \left({\bf r'}\right).$$    
Here $\phi\left({\bf r}\right)$ is the Fourier transform of 
$$\phi({\bf q})\equiv 2^{D-1}\pi Q^{2} / |{\bf q}|^{D-1},$$ as follows
from Poisson equation in three and two-dimensional systems
\cite{ort06} in the presence of the 3D Coulomb interaction and plays the role of the ``effective'' Coulomb
interaction. With these definitions Eq.~\eqref{eq:nq} states 
that the total local electrochemical potential $\mu\left({\bf r}\right)+ v\left({\bf r}\right)$ is
constant.
The latter condition is the generalization to electronic systems of
the Maxwell condition for neutral fluids that is enforced by the
constancy of the local chemical potential across 
different phases. 

We can now use Eq.~\eqref{eq:nq} to eliminate the charge from the
energy and obtain an energy functional that depends only upon $s({\bf q})$:
\begin{eqnarray}
E&=& \sigma \Sigma-\dfrac{\Delta \mu_{c}}{2} s_{0}\left(\overline{n}-n_{c}\right)+\dfrac{V}{2 \kappa}\left(\overline{n}-n_{c}\right)^{2}+  \label{eq:sfunctional}\\ & & -\dfrac{\Delta \mu_{c}^{2}}{8  V}\sum_{{\bf q} \neq 0} \dfrac{s_{{\bf q}} s_{{\bf -q}}}{\left[\kappa^{-1}+2^{D-1}\pi Q^{2}/|{\bf q}|^{D-1}\right]} \nonumber
\end{eqnarray}
where we have dropped an
irrelevant constant. 
 The first term  in Eq.~\eqref{eq:sfunctional} is the surface energy of the sharp interfaces with $\Sigma$ indicating the total surface of the interfaces.
The second and third terms are the ${\bf q}=0$ contribution from the bulk free
energy of the competing phases. The last term comes from the ${\bf q} \neq
0$ contribution of the last three terms of Eq.~\eqref{eq:model} after
 eliminating the charge via Eq.~\eqref{eq:nq}.

Eq.~\eqref{eq:sfunctional} has to be optimized with respect to the
geometry of the domains. Here we restrict to a periodic structure of alternating $s=\pm 1$ stripes. 
This state can be interpreted as a {\it smectic} electronic liquid
phase that possesses orientational order and breaks the translational
symmetry only in one direction. Possibly, fluctuations of the stripe order can
restore the translational symmetry \cite{frad99} thus leading to a
{\it nematic} phase.  
For the smectic stripe state the problem can be solved analytically
(Secs.~\ref{sec:2d},\ref{sec:3d}). Other states with different
geometries like  
circular drops can be treated numerically and will not be
considered in this work. 
\end{section}
\begin{section}{2D systems}
\label{sec:2d}
\begin{figure}[tbp]
\includegraphics[width=8 cm]{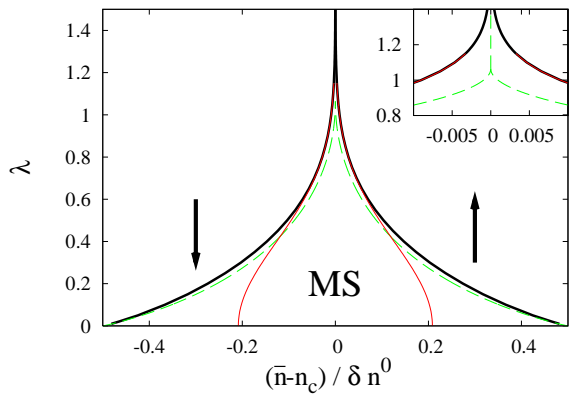}
\caption{(Color online) The phase diagram for the smectic stripe solution in 2D. The central
  region corresponds to the mixed state (MS).  
The thick line is the exact solution of the present model with
finite compressibility,  the thin full line is the $\kappa\rightarrow \infty$ 
limit taking the short distance cutoff of Ref.~\onlinecite{jam05} equal to
the present screening length $l_s$. The dashed line is the UDA. The inset shows an enlargement near the crossing density $n_{c}$ where the UDA predicts a critical value of the frustrating parameter.} 
\label{diagramma2d}
\end{figure}
In this section we consider 2D systems. From Eq.~\eqref{eq:sfunctional}
the energy density ($e\equiv E/V$) of a mixed state  with alternating
$s=1$ stripes of width $2R_{d}$ and $s=-1$ stripes of width
$2(R_{c}-R_{d})$, can be cast as:  
\begin{eqnarray}
  e &= &\dfrac{\sigma}{R_{c}}+\Delta \mu_{c}\left(\dfrac{1}{2}-\nu\right) \left(\overline {n}-n_{c}\right)+\dfrac{ \left(\overline {n}-n_{c}\right)^{2}}{2 \kappa} \label{eq:deltae}  \\  
   &   -&\dfrac{\Delta \mu_{c}^{2}}{2^{D-1} \pi  Q^{2} (2 R_{c})}
\left[u\left(0,\frac{R_{c}}{l_{s}}\right)- u\left(2 \nu ,\frac{R_{c}}{l_{s}}\right) \right] \nonumber
\end{eqnarray} 
where we introduced the two-dimensional
screening length 
$$l_s\equiv (2 \pi Q^2 \kappa)^{-1},$$ 
 and the volume fraction  $\nu\equiv R_{d}/R_{c}$ of the $s=1$ phase.
We also define the Fourier transform of the effective interaction $u_{q}$:
\begin{equation*}
u\left(\frac{x}{R_{c}},\frac{R_{c}}{l_{s}}\right)\equiv
\dfrac{1}{2 R_{c}}\sum_{n\ne 0} e^{i\,q_n \, x} u_{q_n} \hspace{.7cm} q_n\equiv \dfrac{\pi n}{R_{c}} 
\end{equation*} 
with $n$ running over nonzero integers and 

\begin{equation*}
u_{q}\equiv \dfrac{1}{|q|\left(1+l_s|q|\right)}.
\end{equation*}
This effective interaction coincides in this case with the effective 3D screened
Coulomb interaction in two-dimensional electronic systems discussed in
Refs.~\onlinecite{and82} and \onlinecite{ort06}.

Eq.~\eqref{eq:deltae} can be recast in dimensionless form measuring
the energy in units of the  characteristic phase separation energy density gain
$e_0\equiv(\delta n^{0})^{2}/ \kappa$, lengths in units of $l_s$ and
densities in units of $\delta n^0$:
\begin{eqnarray}
 \dfrac{e}{e_0} &=&\lambda^{2}\dfrac{l_s}{ 2 R_c}+\left(\dfrac{1}{2}
-\nu\right) \left(\frac{\overline n-n_{c}}{\delta n^{0}}\right)+\dfrac{1}{2} \left(\frac{\overline n-n_{c}}{\delta n^{0}}\right)^{2} \nonumber \\
&-&\dfrac{l_s}{2 R_c}  
\left[u\left(0,\frac{R_{c}}{l_{s}}\right)- u\left(2 \nu ,\frac{R_{c}}{l_{s}}\right) \right]
\label{eq:deltaedimless}
\end{eqnarray}
Here we defined
a dimensionless coupling:
$$ \lambda^{2}\equiv 4 \pi\sigma Q^2/\Delta\mu_c^{2}$$
that parametrizes the mixed state energy density.
Apart from numerical constants, $\lambda$ coincides with the frustrating parameter defined in  Refs.~\onlinecite{lor01I,ort06} that measures the strength of the frustration due to
LRC interaction and  surface energy with respect to $e_0$.

The dimensionless free energy density, Eq.~\eqref{eq:deltaedimless}, 
 depends upon  $\lambda$, $(\overline n-n_c)/\delta
n^0$, $\nu$ and $ R_{c}/l_{s}$. Minimizing with respect to the volume
 fraction $\nu$ and $R_{c}/l_{s}$ we obtain the phase diagram in the
 $\lambda$-density plane shown in Fig.~\ref{diagramma2d} with the
 thick lines.  

\begin{figure}[tbp]
\includegraphics[width=8 cm]{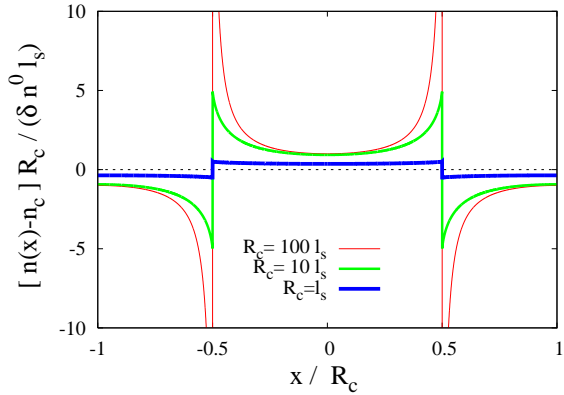}
\caption{(Color online) Charge density modulation for a cut perpendicular to a stripe
  in 2D at $\overline n=n_{c}$, $R_{d}
  = R_{c}/2$ and three different values of the screening length corresponding to $\lambda \sim 0.2$ ($R_{c}=l_{s}$), $\lambda \sim 0.7$ ($R_{c}=10 l_{s}$) and $\lambda \sim 1.1$ ($R_{c}=100 l_{s}$). In the
  case $R_c/l_s \rightarrow \infty$ the charge density diverges at the
  boundary. A finite $l_s$ removes the divergence. }  
\label{rho}
\end{figure}

As in many
 other situations, increasing the frustration, the region of stability
 of the uniform phases is increased. 
 As for the $\kappa=\infty$ case (thin full line in
 Fig.~\ref{diagramma2d}), there is no direct first-order
 transition between the two homogeneous phases no matter how big is
 $\lambda$ \cite{jam05}. Indeed
 the transition line is logarithmically  singular 
at $\overline n=n_c$ and thus for large $\lambda$ the mixed state is enclosed
 in the exponentially narrow range 
 \begin{equation}
   \label{eq:dn}
   |\overline n-n_{c}| <\delta n^0 e^{-\pi \lambda^{2}}.
 \end{equation}
 Clearly, 
observation of the mixed state for large $\lambda$ requires an
 enormously  accurate control of the density $\overline{n}$ which may
 be hard to achieve in practice.
As frustration decreases, the range of densities of the mixed state
grows and converges to the Maxwell construction range when 
$\lambda \rightarrow 0$.

Close to the divergence, the present phase diagram coincides with the
one for infinite compressibility (thin full line in Fig.~\ref{diagramma2d}) provided
one takes the short-distance cutoff of Ref.~\onlinecite{jam05} equal to $l_{s}$.  
This agreement is due to the fact that in the highly frustrated regime
the physics is determined by the slow power-law relaxation
of the charge (Fig.~\ref{rho}) far from the stripe boundary.
 In fact the stripe width is much larger than the screening length and indeed behaves as (c.f. Fig.~\ref{fig:rd2d}) 
\begin{equation}
R_d=\dfrac{R_{c}}{2} \sim l_s e^{\pi\lambda^{2}}
\label{eq:rdlargelambda}
\end{equation}
whereas the finite compressibility of the present model affects the 
charge only in a range of order $l_s$ from the boundary. 
Its effect is to remove the
unphysical divergence of the charge density  at the stripe boundary 
arising when $l_s=0$ (Fig.~\ref{rho}).
. 
On the
contrary, for low $\lambda$ the stripe size is of the
order of the screening length $R_d \sim l_s \lambda $ (Fig.~\ref{fig:rd2d})
and the finite compressibility is relevant.

\begin{figure}[tbp]
\includegraphics[width=8 cm]{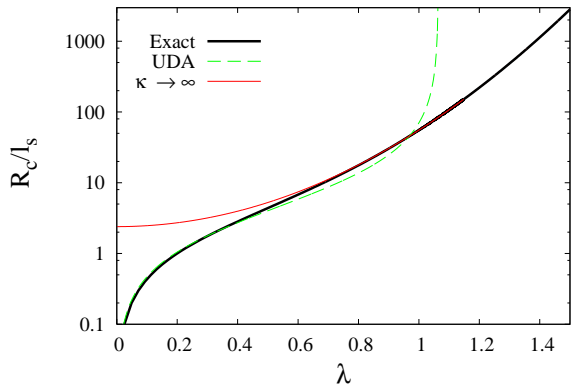}
\caption{(Color online) Behavior of $R_{d}$ (log scale) at the crossing density as
  a function of the frustrating parameter. The thick line is the
  solution of the present model while the thin line is the infinite
  compressibility limit\cite{jam05}. The UDA is the dashed line.}
\label{fig:rd2d}
\end{figure}

For practical applications and for more complicated forms of the energy specifying the homogeneous phases, 
it may be convenient to use a uniform density approximation (UDA) in
which one neglects the spatial variation of the density
inside each inhomogeneity \cite{lor01I,ort06}. 
 The phase diagram in this case is also plotted in Fig.~\ref{diagramma2d} with a
dashed line.
As in Ref.\onlinecite{ort06} the electrostatic energy has been
approximated by taking into account only the self interaction {\it i.e.} we
neglect the Coulomb interaction among different stripes.

\begin{figure}[tbp]
\includegraphics[width=8 cm]{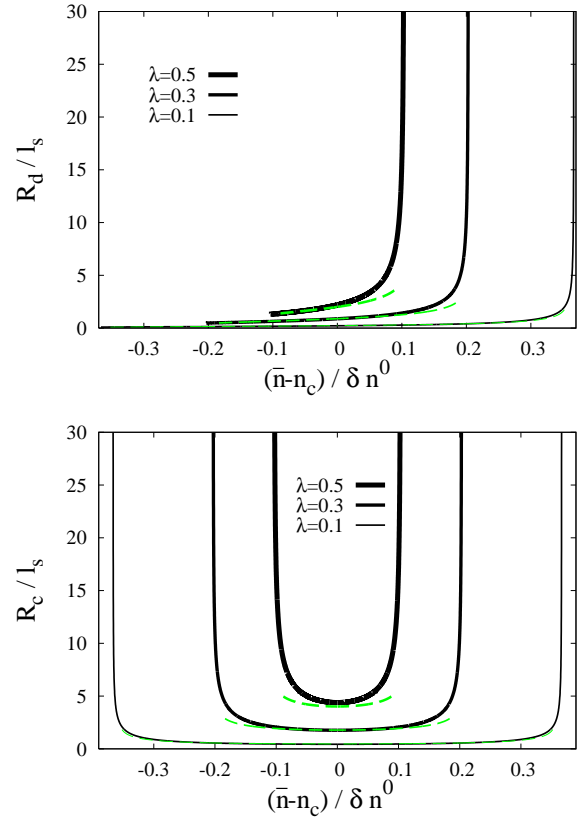}
\caption{(Color online) Behavior of the width (top panel) and
  periodicity (bottom panel) of $\uparrow$ stripes in 2D systems for
  different values of $\lambda<1$. The dashed lines correspond to the
  uniform density approximation whereas the full line is the exact solution. Notice that the $\uparrow$ stripes are the minority phase for $n<n_{c}$ and the majority phase for $n>n_{c}$}
\label{l2d}
\end{figure}

Remarkably the UDA with this crude approximation for the Coulomb
interaction gives very accurate results  
in a wide range of the phase
diagram except around the crossing density, since it misses the
logarithmic singularity and then predicts a critical value $\lambda_{c}$ of the  frustrating parameter above which  the uniform phases would be stable
as in the 3D case \cite{lor01I} and a first order phase transition among them would occur (see the inset of Fig.~\ref{diagramma2d}). Approaching $\lambda_{c}$ from below, the MS disappears with a divergent stripe size (see Fig.~\ref{fig:rd2d}) in contrast with the exact results where the MS persists with a finite stripe size.

\begin{figure}[tbp]
\includegraphics[width=8cm]{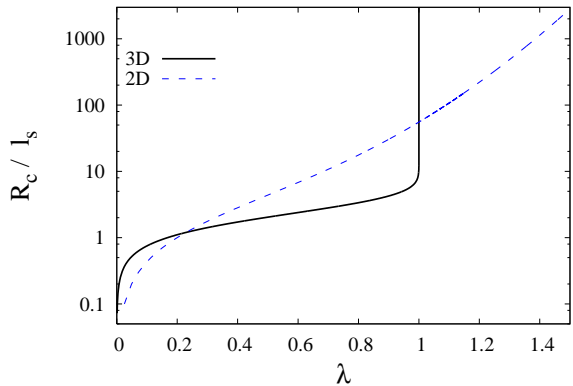}
\caption{(Color online) Behavior of the  period of layers (log scale) vs $\lambda$ at the crossing density in 3D (thick line) and 2D systems (dashed line).}
\label{l3d}
\end{figure}

For $\lambda<\lambda_{c}$,  at the onset of the mixed state the UDA
predicts a first-order transition with the minority phase stripe size
and the stripe periodicity that stay finite (see
Fig.~\ref{l2d}). Taking into account the local dependence of the
electronic density, the jump in the periodicity of the stripes at the
transition is substituted by a square root divergence.
 Thus the uniform-mixed
state transition is characterized by a finite value of the minority phase stripe size
and a divergent periodicity which corresponds to a second-order
transition \cite{note}. Apart from the singularity at the transition
between the uniform phase and the smectic stripe phase, the behavior of 
the mixed state 
for not so large frustration
is very well
represented by the UDA (Fig.~\ref{fig:rd2d},~\ref{l2d}).

For $\lambda>\lambda_{c}$ relaxation of the UDA  produces the logarithmic
singularity which, as stated above, affects in an exponentially narrow
region the phase diagram. 
Furthermore, according to Eq.~\eqref{eq:rdlargelambda}, for large
$\lambda$  the size of the domains can 
quickly reach the size $L$ of the system. Finite $L$ will introduce a 
cutoff to the singularity $$\lambda_{c}^{*}\sim 
\left[(\log{L/l_s})/\pi\right]^{1/2}.$$ 
For $\lambda >\lambda_{c}^{*}$ the
transition will look first order like. 

For $\lambda <\lambda_{c}^* $ the transition is second order like, however 
the effect of the mixed state solution is to produce an
exponentially small rounding of the singularities in thermodynamic quantities 
quite hard to distinguish from a first order jump. 
On the other hand an important difference from a true first order
transition is that the latter shows hysteresis when driven at a finite rate
whereas  here  hysteresis will be absent for  
$\lambda < \lambda_{c}^*$ due to the fact
that the surface energy is effectively negative. For 
$\lambda > \lambda_{c}^*$ instead, hysteresis can appear.

Now we discuss the thermodynamic stability. In Fig.~\ref{energylambda}
we show the energy of the mixed state solutions. For finite
 $\lambda$ the energy vs. density has a negative curvature in the
 mixed state indicating a negative electronic compressibility. 
 For large $\lambda$ the 
electronic compressibility in the mixed state ($|\overline n-n_{c}|/
\delta n^{0}<e^{-\pi \lambda^{2}}$) 
behaves as 
\begin{equation}
  \label{eq:kappa}
\kappa_e^{-1} \sim -\frac{\kappa^{-1}  e^{\pi \lambda^{2}}}{\sqrt{1-[e^{ \pi
    \lambda^{2}}(\overline n-n_{c})/\delta n^{0}]^{2}}} .
\end{equation}
$\kappa_{e}^{-1}$ 
negatively diverges at the uniform-mixed state transition and  
is exponentially large and negative for $\overline n\sim n_c$. 

A negative divergence of the compressibility with a 1/2 critical exponent 
also arises at the uniform-stripe transition for small 
$\lambda$ but with a strength that vanishes when $\lambda\rightarrow 0$.

In the present model the background is assumed to be
incompressible ($k_b=0$) but real systems will have a small background
compressibility $k_b>0$ and the system will become unstable when the
total inverse compressibility $k_b^{-1}+k_e^{-1}<0$.
This will lead to a volume instability \cite{lor01I}, substituting the stripe
transition with a volume collapse transition  and reintroducing
hysteresis \cite{bus05}. 
This is the most likely behavior to be found in real systems specially
for large $\lambda$ where the electronic compressibility is large and
negative in the whole mixed state stability range.

\end{section}
\begin{section}{3D systems}
\label{sec:3d}

In this section we discuss the 3D case.  
The energy of a layered mixed state is given by
Eq.~\eqref{eq:deltae} with $D=3$, 
provided 
the effective interaction $u\left(q\right)$ is given by
$$u\left(q\right)=(1+l_s^{2}q^{2})^{-1},$$ and the screening length $l_{s}$ is substituted  by 
 the usual 3D expression,  $$l_s=(4\pi Q^2 \kappa)^{-1/2}.$$ 
Notice that the standard  3D screened Coulomb interaction is given by
$u\left(q\right)/q^{2}$.

The different nature of the screening
in two- and three- dimensional systems affects strongly the 
properties of the mixed state. At the crossing density, the exact expression
for the energy density of a layered state is minimized for
$\nu=R_{d}/R_{c}=1/2$ and takes the following simple form:
\begin{equation}
e=\frac{\sigma}{R_{c}}
\left[1- \lambda^{-3/2}\tanh{\frac {R_{c}}{2 l_s}}\right]
\label{eq:3dene}
\end{equation}
where $\lambda$ is the 3D frustrating parameter: \cite{lor01I} 
$$\lambda=4
\left[( \pi Q^{2} \sigma^{2})/({\kappa \Delta
    \mu_{c}^{4}})\right]^{1/3}.$$ 

The energy density of the uniform state is given by $e=0$, hence 
for a layered state to be possible the
term in the brackets in Eq.~\eqref{eq:3dene} must
become negative.
This condition is equivalent to the existence of a critical
frustrating parameter $\lambda_{c}= 1$ 
as it was derived in Ref.~\onlinecite{lor01I} within the UDA and now showed in an exactly
solvable model. 

Minimizing Eq.~\eqref{eq:3dene} respect to $R_c$ one finds that $R_c$
is smaller than a few screening lengths except in the unphysical
case in which $\lambda$ is fine-tuned exponentially close to
$\lambda_c$. This is in contrast with the 2D case
where $R_c$ is unbounded for a generic large $\lambda$ (Fig.~\ref{l3d}).

\begin{figure}[tbp]
\includegraphics[width=8cm]{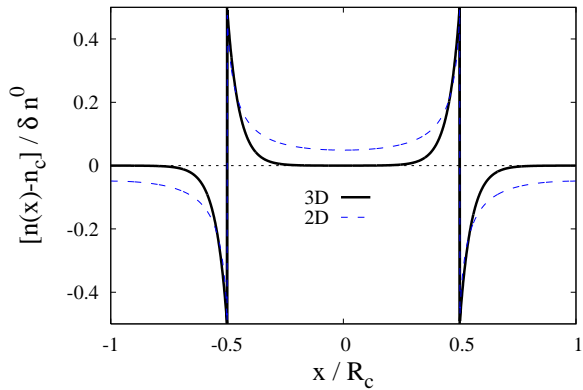}
\caption{(Color online) Charge density modulation for a cut perpendicular to the
  layers in 3D systems (thick line) and 2D systems (dashed line) at
   $\overline{n}=n_{c}$ and a layer period $R_{c}=20 l_{s}$.} 
\label{rho3d}
\end{figure}

This result is due to the different screening effect in
three-dimensions with respect  to the 2D case. To illustrate this
difference we plot in
Fig.~\ref{rho3d}, the charge density in 2D and 3D in a cut
perpendicular to the stripe/layer for $R_c=20 l_s$. 
In 3D the
charge density decays exponentially from the interface over a distance
of the order of the screening length whereas  in 2D it decays as a power
law. As emphasized in Ref.~\onlinecite{lor01I} the phase separation
energy gain stems from the regions where the local density $n(x)$ is  
significantly different from the global $\overline{n}$ value. In 2D this is
generically fulfilled and thus phase separation is favored. In 3D
instead,  the two densities are
substantially different only in a region of width $l_s$ around the
interface. For $\lambda<\lambda_c$ the phase separation energy gain
from these regions compensates the surface and Coulomb energy cost and
makes inhomogeneities possible. Instead the central region in the 
3D case of Fig.~\ref{rho3d} produces an exponentially small phase 
separation energy gain and
therefore one never finds inhomogeneities with  $R_c>> l_s$, as in the
figure, unless $\lambda$ is exponentially close to $\lambda_c$. Indeed 
one can check from Fig.~\ref{l3d} that this is the case for the
present value of $R_c/l_s$.

\begin{figure}[tbp]
\includegraphics[width=8cm]{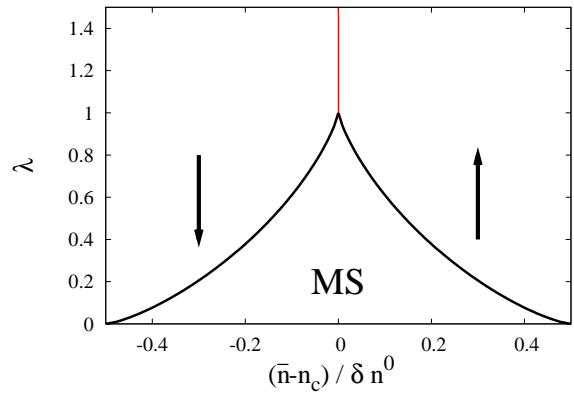}
\caption{ (Color online) The phase diagram in the
  $\lambda-(\overline{n}-n_{c})/ \delta n^{0}$ plane for
  three-dimensional systems. The thick lines correspond to the
  second-order transitions from the homogeneous phase to the smectic stripe phase characterized by a divergent periodicity and a finite minority
  phase layer
  size. The thin line represents the first order transition between the two homogeneous phases that occurs for $\lambda>\lambda_{c}$.}
\label{phasediagram3d}
\end{figure}

The fact that strong variations of the density as the one pictured in
Fig.~\ref{rho3d} practically never occur in 3D makes the UDA 
reliable everywhere in the 3D phase diagram.  

The exact 3D phase diagram for the layered state in the $\lambda$-density
plane is shown in Fig.~\ref{phasediagram3d}. 
As expected it is very similar to the one obtained within the UDA in
Refs.~\onlinecite{lor01I,lor02}. The only difference is that in the
UDA the uniform-inhomogeneous transition resulted weakly first order
whereas here it is second-order like due to a divergence of $R_c/R_d$
as in 2D.

The main difference with the 2D case is that the two second order
transition lines from the homogeneous phase to the mixed state touch each other
at $\lambda=\lambda_{c}$.  For $\lambda>\lambda_{c}$ the systems has no intermediate state
between the uniform phases and a direct first-order transition appears
among them. In this case the energy density as a function of the global density is given by the 
lowest energy branch among 
the two parabolas shown in Fig.~\ref{energylambda}. This has a cusp
singularity at $n_c$ which produces a Dirac function like negative
divergence of the compressibility. This leads to a
volume instability  also in this case\cite{lor01I,bus05}.

\end{section}

\begin{section}{Conclusions}
\label{sec:conclusions}
In this work we have solved a model of phase separation frustrated by
the LRC interaction while considering screening effects and the charge
relaxation inside the domains in two and three dimensional systems.

 In 2D and for high
frustration the finite bare compressibility can be neglected. Its effect
can be taken up by a short distance cutoff.  In this regime,
however,  the mixed state requires very stringent conditions to be
observed: {\it i}) it requires an enormously  accurate
control of the density since it appears in an exponentially narrow region
around the crossing density, {\it ii}) the size of the domains grows exponentially with the frustration and may easily reach the size of the system , {\it iii})  the mixed state
is characterized by an exponentially large negative compressibility
which induces a volume instability when the rigidity of the
background is not sufficient to stabilize the system altogether.  

An incompressible background may seem unphysical according to the 
above discussion. However in some cases a sufficient
separation of energy scales may avoid a volume instability
if singularities in the compressibility are rounded by extrinsic
effects. For example in ruthanates and in the 2D electron gas the
relevant electronic phenomena occurs at
temperatures below a tenth of a Kelvin to be 
compared with melting temperatures of
the material of the order of hundreds of Kelvin.   
Even when a volume instability may be avoided, the strongly frustrated mixed state may appear somewhat academic due to the
stringent conditions on the density. An important physical, however, 
consequence is that the transition will look first order like but
without hysteresis as discussed in Sec.~\ref{sec:2d}.  In the case of
the ruthanates there is the additional advantage that the control
parameter is the magnetic field which allows for considerable fine-tuning.  

For low frustration the 
 finite bare compressibility of the phases [the third term in
 Eq.~\eqref{eq:model}] cannot be neglected. The mixed phase
 compressibility gets strongly renormalized. We have discussed 
 its behavior in 2D [Eq.~\eqref{eq:kappa}] 
which in principle can be directly measured by capacitive techniques\cite{eis94}. We have also shown that neglecting the spatial variation of the
charge inside the domains, one obtains very accurate results at low
 frustration. From  our results and Figs.~\ref{diagramma2d},\ref{fig:rd2d} it is clear
 that the limit $\kappa \rightarrow \infty $ and the UDA can be seen
 as the strong  and the weak frustration  approximation respectively for the full
 model Eq.~\eqref{eq:model}.

In 3D systems the bare compressibility term 
is essential to obtain the mixed state with sharp
interfaces as studied here and the UDA becomes reliable everywhere in
the phase diagram.   

The different role of the screening in 2D and 3D can be
interpreted by noticing that the electronic charge density behavior
inside the domains is dramatically different in 3D with respect to
2D. In the latter case, in fact, the charge density inside one stripe
decays with a power-law from the interface  
and the local electronic charge density differs from the background
density (i.e. $n\left({\bf x}\right) \neq \overline{n}$) over the
entire stripe width. Thus domains gain phase separation energy in all
the region and there is no limitation for their size.   
This also explains why the mixed state can appear independently on how strong the frustration is.

 On the contrary, in 3D systems
 the charge density decays exponentially. As a consequence, 
the system is forced to satisfy a ``maximum size
rule''\cite{lor01I,lor02} that states that for generic parameters
  the inhomogeneities cannot have all linear dimensions much larger
 than the 3D screening length $l_s^{3D}$.  
This allows for arbitrary large
inhomogeneities in 2D since one of the dimensions is already smaller
than $l_s^{3D}$ as indeed found (Fig.~\ref{l2d}). 

The frustrating parameter can be written as
\begin{equation*}
\lambda=\left[2^{D-1} \dfrac{l_{d}}{l_{s}}\right]^{1-\frac{1}{D}}
\end{equation*}
Here $l_{d}$ represents the size that inhomogeneities should have for
the surface energy cost, $\sigma / l_{d}$, be of the same order as the
phase separation energy gain, $e_{0}=\left(\delta n^{0}\right)^{2} /
\kappa$.  

To make a rough estimate of $\lambda$ we go back to the lattice and take the bare
compressibility as an inverse bandwidth $\kappa=\left[2 D t\right]^{-1}$ with $t$ an
interatomic hopping integral. Defining a nearest neighbor Coulomb
interaction $V\equiv Q^2/a$
 one obtains:

\begin{eqnarray*}
\lambda^2&=&\frac {\pi}2  \frac{V J }{t^2 (\delta n_0)^2}  \hspace{.7cm} (D=2) \\ 
\lambda^3&=&\frac {32\pi}{27}  \frac{ V J^2}{t^3 (\delta n_0)^4}  \hspace{.7cm} (D=3)  
\end{eqnarray*}
If one neglects the logarithmic singularity in 2D, inhomogeneities
require $\lambda \lesssim 1$. In general we expect the scale $J$ which
may be due to magnetism or other low energy phenomena to satisfy
$J<t$. According to our definitions $\delta n_0$ is the difference of the Maxwell construction
densities (measured as number of particles per site). 
Typically we expect weak phase separation tendency 
so $\delta n_0<<1$. Comparing with the 2D case, 
the 3D case has an extra factor $J/t$ that makes the mixed state more
favorable and an extra factor $\delta n_0^{-2}$ that makes the mixed
state less favorable. A more precise analysis requires
microscopic modeling of specific situations and will be presented
elsewhere.

\end{section}

\end{document}